# Evidence for a long-lived superheavy nucleus with atomic mass number A = 292 and atomic number Z ≅ 122 in natural Th

A. Marinov[1], I. Rodushkin[2], D. Kolb[3], A. Pape[4], Y. Kashiv[1], R. Brandt[5], R.V. Gentry[6] &

H.W. Miller[7]

[1]*Racah Institute of Physics, The Hebrew University of Jerusalem, Jerusalem 91904, Israel*

[2]*Analytica AB, Aurorum 10, S-977 75 Luleå, Sweden*

[3]*Department of Physics, University GH Kassel, 34109 Kassel, Germany*

[4]*IPHC-UMR7178, IN2P3-CNRS/ULP, BP 28, F-67037 Strasbourg cedex 2, France*

[5]*Kernchemie, Philipps University, 35041 Marburg, Germany*

[6]*Earth Science Associates, P.O. Box 12067, Knoxville, TN 37912-0067, USA*

[7]*P.O. Box 1092, Boulder, CO 80306-1092, USA*

Evidence for the existence of a superheavy nucleus with atomic mass number A=292 and abundance $(1-10) \times 10^{-12}$ relative to $^{232}$Th has been found in a study of natural Th using inductively coupled plasma-sector field mass spectrometry. The measured mass matches the predictions[1,2] for the mass of an isotope with atomic number Z=122 or a nearby element. Its estimated half-life of $t_{1/2} \geq 10^8$ y suggests that a long-lived isomeric state exists in this isotope. The possibility that it might belong to a new class of long-lived high spin super- and hyperdeformed isomeric states is discussed.[3-6]

The question "how heavy can a nucleus be" is a fundamental problem in nuclear physics.[7] Experimentally, elements up to Z=118 have been produced synthetically by heavy-ion reactions,[8,9] with the half-lives of the Z=106 to 118 nuclei ranging from a few minutes to about a millisecond. However, in a recent study of natural Th substances,[10] long-lived isomeric states with estimated half-lives $t_{1/2} \geq 10^8$ y, 16 to 22 orders of magnitude longer than their corresponding ground states (g.s.), have been observed in the neutron-deficient $^{211,213,217,218}$Th nuclei. It has been hypothesized[10] that they might belong to the recently discovered[3-6] class of long-lived high spin isomeric states in the second minimum (superdeformed (SD) minimum)[11,12] and/or in the third minimum (hyperdeformed (HD) minimum)[13-15] of the nuclear potential energy when plotted as a function of deformation. This result motivated us to search for long-lived superactinide elements in natural materials. According to the extended periodic table,[16] the superactinides are classified as homologues of the actinides, with elements 122 and 124 placed as eka-Th and eka-U, respectively. As a working hypothesis, it is reasonable to assume that if any of the superactinide elements exists in nature they might be carried along with Th and/or U substances. Here we report evidence for the existence of a nucleus with atomic mass number 292 that was found in natural Th with an estimated half-life of $\geq 10^8$ y and an abundance of $(1-10) \times 10^{-12}$ relative to $^{232}$Th. Its measured mass matches the predictions[1,2] for the mass of an isotope with atomic number Z=122 and also some isobars of neighbouring elements. Based on theoretical chemical predictions[17] it is argued that it is probably element 122. Preliminary results of this work have been presented previously.[18]

The mass $M_A$ of an atom is equal to:

$$M_A = Z \times M_H + N \times M_n - BE \qquad (1)$$

where $M_H$ and $M_n$ are the masses of the hydrogen atom and the neutron, respectively, and BE is the binding energy of the nucleus. The binding energy per nucleon (BE/u) of stable nuclei has a broad maximum around A ≈ 60, with a value of 8.7 MeV/u, which falls monotonically to about 7.6 MeV/u at Th and U.[19] The predicted BE/u values for the superactinide nuclei are around 6.9 MeV/u.[1,2,20,21] Therefore, masses of superactinide isotopes are higher than, and resolvable from, the masses of all molecules with the same mass number, except for multihydrogen-containing molecules. This is seen in Fig. 1, where the systematic behaviour of the masses[22] of various M=292 species, from the symmetric combination $^{146}$Nd$_2$ to Pb-, Th- and U-based molecules to the predicted[1,2] masses of the $^{292}$122 and $^{292}$124 nuclei are displayed. The masses of the very neutron-rich $^{292}$Th and $^{292}$U nuclei are predicted to be 292.399 and 292.362 u,[20] values well above the expected masses of the $^{292}$122 and $^{292}$124 nuclei, that are 292.243 and 292.264 u, repectively.[1,2] (The calculated masses of these two nuclei in Refs.[1,2] agree to within 0.001 u.) Thus, as was demonstrated before,[23] accurate mass measurements are an effective tool in searching for naturally occurring superheavy elements.

In principle, natural minerals like monazite, which is the usual source material for Th, would be the most promising materials to study. However, background was the main obstacle when looking for isotopes with relative abundance of $(1-10) \times 10^{-11}$ in natural materials.[10] Therefore, purified natural Th was used in our measurements.

In the present work, we performed accurate mass measurements for masses 287 to 294 in Th solutions. Evidence was obtained for the existence of an isotope with a mass that matches the predictions for atomic mass number 292 and Z around 122. Here we describe this observation.

The experimental procedure was similar to that described earlier.[10] Inductively coupled plasma-sector field mass spectrometry (ICP-SFMS) was used for the experiments. The ICP-SFMS is an Element2 (Finnigan, Thermo-Electron, Bremen, Germany). In this instrument, a solution of the material to be studied is introduced into a high temperature (6000-8000 K) plasma source. At these temperatures, predominantly atomic species are present. Molecular ions are produced after the source, mainly by interaction with oxygen and hydrogen ions from the solution. The predefined medium resolution mode, $m/\Delta m = 4000$ (10% valley definition), was used throughout to separate atomic ions from molecules with the same mass number. The sensitivity-enhanced set-up of the instrument was similar to that described in Ref.[24] This set-up provided sensitivity for $^{232}$Th in this resolution mode of up to $2 \times 10^8$ counts $s^{-1} mg^{-1} l^{-1}$. The sample uptake rate was 60-80 $\mu l\ min^{-1}$. Methane gas was added to the plasma to decrease the formation of molecular ions.[25] Oxide and hydride formation (monitored as $UO^+/U^+$ and $UH^+/U^+$ intensity ratios) were approximately 0.04 and $1 \times 10^{-5}$, respectively. In its standard mode of operation, the measured masses in the ICP-SFMS are limited to M≤270. In order to measure higher masses, the accelerating voltage was reduced from 8 to 7 kV. The machine was calibrated using the $^{230}$Th, $^{235}$U, $^{238}$U, $^{238}U^{16}O$ (M=254) and $^{238}U^{40}Ar^{16}O$ (M=294) molecules.* Then a new calibration was performed using the $^{232}$Th$^{40}$Ar$^{16}$O (M=288) and $^{238}$U$^{40}$Ar$^{16}$O molecules. This calibration was checked regularly every half to one hour of the experiments and the data were corrected accordingly.

1000 mg l$^{-1}$ purified Th solutions were purchased from two companies. The solutions were analysed in three sessions: February 25 (run I), May 25 (run II), and July 5 (run III), 2007. A range of about 0.42 u, divided into 60 channels, was scanned in each measured spectrum in the first two runs, and a range of about 0.2 u, divided into 30 channels, was scanned in the third run.

During the first session, masses from 287 to 294 were analysed using the Th solutions with an integration time of 1 s channel$^{-1}$, and these measurements were made three to five times each. In one spectrum, two events were observed with a solution produced by Customer Grade (from LGC Promochem AB, Borås, Sweden) at a mass that fits the predictions for atomic mass number 292 and a Z value around 122. This mass region was again studied in runs II and III with the same solution.

Complete elemental screening was performed on the Th solution to assess the impurity concentration levels. The concentrations of certain trace elements that could potentially give rise to spectrally interfering molecular species, expressed as ppm of the Th concentration, are as follows:

U 80, Bi <0.1, Pb 0.2, Hg <0.1, Au <0.1, Os <0.1, Hf <0.1, Dy 0.7, Nd 3, Ce 0.7, La 0.4, Ba 0.2, Cs 0.2, I 1.0, Te <0.1, Fe 7, Ni 3, Ca 50, K 1000, B 3, Be <0.1

Instrumental sensitivity varied between runs as a result of matrix effects caused by the introduction of highly concentrated solutions into the ICP source. During the first run, the Customer Grade Th solution, diluted to 50 mg Th l$^{-1}$ of 0.7 M HNO$_3$, was studied. Figure 2 shows the results in the mass region 288 where the peak of the $^{232}$Th$^{40}$Ar$^{16}$O molecule is observed. The centre of mass (c.m.) of this peak is shifted from the known mass of this molecule[22] by 0.014 u. The full width at half maximum of this peak is about 0.040 u.

---

* Ar is the carrier gas in the ICP-SFMS.

The results obtained in run I for the mass region of 292 are displayed in Fig. 3. Figure 3(a) shows a spectrum where the integration time per channel was 1 s. The two events seen in this spectrum are at mass 292.232 u. (A correction of -0.014 u due to the calibration shift mentioned above was applied in deducing the mass.) The observed mass is close to the predicted values of the $^{292}$122 and $^{292}$124 nuclei as indicated by the two arrows in this figure.

Figure 3(b) shows the result of a sum of five spectra, which includes the spectrum seen in Fig. 3(a). As can be seen, only two background events at lower mass were added in the additional four spectra.

In run II, mass 292 was scanned 60 times with the same 50 mg l$^{-1}$ Th solution as in run I. Figure 4(a) displays the results of three spectra out of the 60 measured. In addition to the $^{238}$U$^{40}$Ar$^{14}$N peak, four events are seen in the vicinity of the predicted masses of the isotopes $^{292}$122 and $^{292}$124. Figure 4(b) presents the sum of the 60 spectra. Compared with Fig. 4(a) two events were added in the region from 292.130 to 292.320 u in the additional 57 spectra. This shows that the four events around mass 292.260 u are not background. (The reason that this group has been observed in only a few spectra out of the 60 measured is probably because of its low abundance. For a solution concentration of 50 mg l$^{-1}$, an abundance of 1x10$^{-10}$ will result in 1 count s$^{-1}$ on average. A lower abundance of about 5x10$^{-12}$ will result in one event every 20 s, or about one measurement in 20.) Figure 4(c) shows the sum of the 60 spectra measured with a blank solution of 0.7 M HNO$_3$. The c.m. of the high mass peak seen in Figs. 4(a) and 4(b) is 292.272 u, taking into account a correction of 0.013 u deduced from the measured shift of the $^{238}$U$^{40}$Ar$^{14}$N peak seen in Fig. 4(a).

In run III the concentration of the Th was increased to 80 mg l$^{-1}$. In this run we focused on the region between 292.130 and 292.320 u (denoted by the small upward arrows on the X-scale in Fig. 4(b)), which was divided to 30 channels and was scanned 200 times. Each spectrum was measured with an integration time of 1 s channel$^{-1}$. The calibration was determined before and after every 100 spectra using the $^{232}$Th$^{40}$Ar$^{16}$O (M=288) and $^{238}$U$^{40}$Ar$^{16}$O (M=294) peaks. The sum spectrum of the 200 scans is seen in Fig. 5(a). An equivalent spectrum but with a blank solution is seen in Fig. 5(b). An accumulation of 36 events is seen in Fig. 5(a) at a c.m. position of 292.263 u. The background in the same region in Fig. 5(b) is 7 counts. The background in an equivalent region from 292.154 to 292.226 u in Fig. 5(a) is 14 counts. Taking into account the average of these two backgrounds, the net number of counts is 25±7. The error was estimated according to the formula $\sigma = \sqrt{(Nt + Nb)}$, where $Nt$ is the total number of counts and $Nb$ is the number of background counts.[26]

The weighted average of the mass of the newly observed peak seen in the three runs is 292.262±0.030 u. It is estimated that the concentration of the species responsible for this peak is $(1-10) \times 10^{-12}$ of $^{232}$Th, or about $(1-10) \times 10^{-16}$ of the solution.

Statistical analysis, done in the same way as described earlier,[10] shows that the probability that the observed 292.262 u peak is due to an accidental concentration of events is about $3 \times 10^{-5}$, for each of the three measurements separately.

We were unable to match the signals of the suspected superactinide isotope with any molecular ion. As described in the introduction and displayed in Fig. 1, because of binding energy, masses of molecules are lower than the mass of the observed peak. As already mentioned, the mass of the very neutron-rich $^{292}$Th nucleus is predicted to be 292.399 u,[20] well above the measured peak.

Another possibility that has to be considered is the potential presence of hydrocarbon-based molecular ions from pump oil. However, no contamination from pump oil was seen in the spectra of the blank.

At the same time, the measured mass fits the predictions for the atomic masses of superactinide isotopes with A=292 and Z values from 121 to 126,[1,2] which vary monotonically from 292.236 to 292.291 u, respectively. The prediction for isotope A=292 of element 122 (eka-Th) is 292.243 u. The atomic configuration of Th is $6d_{3/2}^2 7s^2$ and its chemical separation is based on its stable $4^+$ oxidation state. The predicted configuration of element 122 is $8s^2 7d_{3/2} 8p_{1/2}$.[17] In spite of the different configuration, element 122 is also expected to form a stable $4^+$ oxidation state. Element 121 has only three electrons outside the expected filled shells of element 118 (eka-Rn). It seems unlikely that it would form a stable $4^+$ oxidation state. Elements above Z=122 have more electrons outside the closed shells, but their configurations have not been accurately calculated. One can assume that they might form a $4^+$ state but also higher oxidation states. If element 122 exists in nature with Th, then it would be reasonable to assume that it followed the chemical separation of Th and showed up in our measurements. However, the possibility that the observed A=292 isotope might belong to an element of somewhat higher Z cannot be excluded.

The predicted half-lives of the normal g.s. of nuclei around $^{292}$122 are of the order of $10^{-8}$ s.[21] This suggests that the observed events are due to a long-lived isomeric state in $^{292}$122 or in a nearby isobar. If its initial terrestrial concentration was similar to that of $^{232}$Th, then the lower limit on its half-life would be about $10^8$ y, or otherwise it would have decayed away.

The character of the observed isomer has not been measured directly. It cannot be related to the actinide fission isomers, since their lifetimes are in the ns to ms region. As mentioned in Ref.[10], it is also not reasonable to assume that it is a normal high spin K-type isomer,[27] because the lifetimes of all the known K-isomeric states in neutron-deficient nuclei with Z≥84 are not longer than several minutes. It could be a high spin isomer near doubly closed shells. In addition to the recurrent predictions that Z=114 and/or 126 and N=184 could be the next proton and neutron closed shells, respectively,[7] the isotope $^{292}$120 has also been predicted recently to be a doubly closed shell nucleus.[28,29] However, the $Q_\alpha$ value for the $^{292}$122 nucleus should be about 14 MeV.[1,2] It is estimated that for such high energy, a half-life of about $10^8$ y would be obtained by a $\Delta L_\alpha \approx 30\hbar$. Such a high spin difference seems unlikely. One can hypothesize, as we did in the case of the Th isomers,[10] that the proposed state is an aligned high spin SD or HD state.[3-6]

High spin states in general and such states in the SD and HD minima in particular are made preferentially by heavy ion reactions.[6,30] If the observed state in A=292 turns out to be of the high spin SD or HD type, then heavy ion reactions could be involved in its nucleosynthesis.

In summary, mass spectral evidence has been obtained for the existence of a long-lived superheavy isotope with an atomic mass number of 292 and $t_{1/2} \geq 10^8$ y. Based on predicted chemical properties of element 122, it is probable that the isotope is $^{292}$122, but a somewhat higher Z cannot absolutely be excluded. Because of its long lifetime, it is deduced that a long-lived isomeric state rather than the normal g.s. was observed at A=292. The hypothesis that it is a high spin SD or HD isomeric state is discussed.

1. Koura, H., Tachibana, T., Uno, M. & Yamada, M. KTUY mass formula. *Prog. Theor. Phys.* **113,** 305 (2005); KTUY05\_246512np.pdf.

2. Liran, S., Marinov, A. & Zeldes, N. Semiempirical shell model masses with magic number *Z*=126 for superheavy elements. *Phys. Rev. C* **62,** 047301 (2000) 4 pages; arXiv:nucl-th/0102055.

3. Marinov, A., Gelberg, S. & Kolb, D. Discovery of strongly enhanced low energy alpha decay of a long-lived isomeric state obtained in $^{16}$O + $^{197}$Au reaction at 80 MeV. *Mod. Phys. Lett.* **A11,** 861-869 (1996).

4. Marinov, A., Gelberg, S. & Kolb, D. Evidence for long-lived proton decay not far from the stability valley produced by the $^{16}$O + $^{197}$Au reaction at 80 MeV. *Mod. Phys. Lett.* **A11,** 949-956 (1996).

5. Marinov, A., Gelberg, S. & Kolb, D. Discovery of long-lived shape isomeric states which decay by strongly retarded high-energy particle radioactivity. *Int. J. Mod. Phys.* **E10,** 185-208 (2001).

6.  Marinov, A., Gelberg, S., Kolb, D. & Weil, J. L. Strongly enhanced low energy alpha-particle decay in heavy actinide nuclei and long-lived superdeformed and hyperdeformed isomeric states. *Int. J. Mod. Phys.* **E10,** 209-236 (2001).

7.  Kumar, K. *Superheavy Elements* (Adam Hilger, Bristol and New York, 1989).

8.  Hofmann, S. & Münzenberg, G. The discovery of the heaviest elements. *Rev. Mod. Phys.* **72,** 733-767 (2000).

9.  Ogannesian, Yu. Heaviest nuclei from $^{48}$Ca-induced reactions. *J. Phys. G: Nucl. Part. Phys.* **34,** R165-R242 (2007).

10. Marinov, A., Rodushkin, I., Kashiv, Y., Halicz, L., Segal, I., Pape, A., Gentry, R.V., Miller, H.W., Kolb, D. & Brandt, R. Existence of long-lived isomeric states in naturally-occurring neutron-deficient Th isotopes. *Phys. Rev. C* **76,** 021303(R) (2007) 5 pages.

11. Polikanov, S.M. et al. Spontaneous fission with an anomalously short period I. *Sov. Phys. JETP* **15,** 1016 (1962).

12. Twin, P.J. et al. Observation of a discrete-line superdeformed band up to 60ℏ in $^{152}$Dy. *Phys. Rev. Lett.* **57,** 811-814 (1986).

13. Möller, P., Nilsson, S.G. &. Sheline, R.K. Octupole deformations in the nuclei beyond $^{208}$Pb. *Phys. Lett.* **B40,** 329-332 (1972).

14. Ćwiok, S., Nazarewicz, W., Saladin, J. X., Ociennik, W. P. & Johnson A. Hyperdeformations and clustering in the actinide nuclei. *Phys. Lett.* **B322,** 304-310 (1994).

15. Krasznahorkay, A. et al. Experimental evidence for hyperdeformed states in U isotopes. *Phys. Rev. Lett.* **80,** 2073-2076 (1998).

16. Seaborg, G. T. Elements beyond 100, present status and future prospects. *Ann. Rev. Nucl. Sci.* **18,** 53-152 (1968).

17. Eliav, E., Landau, A., Ishikawa, Y. & Kaldor, U. Electronic structure of eka-thorium (element 122) compared with thorium. *J. Phys. B: At. Mol. Opt. Phys.* **35,** 1693-1700 (2002).

18. Marinov, A., Rodushkin, I., Kolb, D., Pape, A., Kashiv, Y., Brandt, R., Gentry, R.V. & Miller, H. W. Possible existence of a superactinide nucleus with atomic mass number 292 in natural Th. *Third Int. Conf. on the Chemistry and Physics of the Transactinide Elements,* Davos, Switzerland, (2007) p. 65.

19. Evans, R.D. *The Atomic Nucleus* (McGraw-Hill, New York, 1955).

20. Möller, P., Nix, J. R., Myers, W. D. & Swiatecki, W.J. Nuclear ground-state masses and deformations. *At. Data Nucl. Data Tables* **59,** 185-381 (1995).

21. Möller, P., Nix, J. R. & Kratz, K.-L. Nuclear properties for astrophysical and radioactive-ion-beam applications. *At. Data Nucl. Data Tables* **66,** 131-343 (1997).

22. Audi, G., Wapstra, A. H. & Thibault, C. The AME2003 atomic mass evaluation (II). Tables, graphs and references. *Nucl. Phys.* **A729**, 337-676 (2003).

23. Marinov, A., Rodushkin, I., Pape, A., Kashiv, Y., Kolb, D., Brandt, R., Gentry, R.V., Miller, H.W., Halicz, L. & Segal, I. Existence of long-lived isotopes of a superheavy element in natural Au. *Third Int. Conf. on the Chemistry and Physics of the Transactinide Elements,* Davos, Switzerland, (2007) p. 66; *arXiv:nucl-ex/0702051*.

24. Rodushkin, I., Engström, E., Stenberg, A. & Baxter, D. C. Determination of low-abundance elements at ultra-trace levels in urine and serum by inductively coupled plasma-sector field mass spectrometry. *Anal. Bioanal. Chem.* **380,** 247-257 (2004).

25. Rodushkin, I., Nordlund, P., Engström, E. & Baxter, D. C. Improved multi-elemental analyses by inductively coupled plasma-sector field mass spectrometry through methane addition to the plasma. *J. Anal. At. Spectrom.* **20**, 1250-1255 (2005).

26. Bevington, P. R. *Data Reduction and Error Analysis for the Physical Sciences* (McGraw-Hill, New York, 1969).

27. Walker, P. M. & Carroll, J. J. Ups and downs of nuclear isomers. *Physics Today* **58**, 39-44 (2005).

28. Rutz, K., Bender, M., Bürvenich, T., Schilling, T., Reinhard, P.-G., Maruhn, J. A. & Greiner, W. Superheavy nuclei in self-consistent nuclear calculations. *Phys. Rev. C* **56,** 238-243 (1997).

29. Kruppa, A. T., Bender, M., Nazarewicz, W., Reinhard, P.-G., Vertse, T. & Ćwiok, S. Shell corrections of superheavy nuclei in self-consistent calculations. *Phys. Rev. C* **61**, 034313 (2000) 13 pages.

30. Marinov, A., Gelberg, S., Kolb, D., Brandt, R. & Pape, A. New outlook on the possible existence of superheavy elements in nature. *Phys. At. Nucl.* **66,** 1137-1145 (2003).

**Acknowledgements**

We appreciate valuable discussions and the help of N. Zeldes, L. Halicz, I. Segal, F. Oberli, U. Kaldor and E. Aliav.

Correspondence and requests for materials should be addressed to A. Marinov (email: *marinov@vms.huji.ac.il*).

**Figure captions**

**Figure 1.** Representation of the systematic behaviour of the masses of various M=292 species, from the symmetric combination $^{146}Nd_2$ to Pb-, Th- and U-based molecules,[22] to the predicted[1,2] masses of $^{292}122$ and $^{292}124$ nuclei.

**Figure 2.** Results of mass measurements for mass region 288. The c.m. of the observed peak is 288.009 u and the known mass[22] of $^{232}Th^{40}Ar^{16}O$ is 287.995 u.

**Figure 3.** Results of measurements for mass region 292 obtained in the first run. A single spectrum out of five measured is shown in (a). The sum of the five spectra is displayed in (b).

The arrows indicate the positions of the predicted[1,2] masses of the $^{292}$122 [$^{292}$122(pred.)] and $^{292}$124 [$^{292}$124(pred.)] isotopes.

**Figure 4.** Results of measurements for mass region 292 obtained in the second run. The sum of three spectra out of the 60 measured is shown in (a), and the sum of the 60 spectra is shown in (b). The two rightmost arrows indicate the positions of the predicted[1,2] masses of the $^{292}$122 [$^{292}$122(pred.)] and $^{292}$124 [$^{292}$124(pred.)] isotopes. The left-hand arrows (in (a) and (c)) indicate the known position[22] of $^{238}$U$^{40}$Ar$^{14}$N. The c.m. of this peak is 292.003 u and its known mass is 292.016 u. The small upward pointing arrows on the X-scale in (b) at masses 292.130 and 292.320 u indicate the limits chosen for the third run. The sum of 60 spectra of the blank solution is shown in (c).

**Figure 5.** Results of measurements for mass region 292 obtained in the third run. The sum of 200 spectra in the limited mass region of 292.130 to 292.320 u (corresponding to the upward arrows on the X-scale in Fig. 4(b)) is shown in (a). Figure 5(b) shows the sum of 200 blank solution spectra. The arrows indicate the positions of the predicted[1,2] masses of the $^{292}$122 [$^{292}$122(pred.)] and $^{292}$124 [$^{292}$124(pred.)] isotopes.

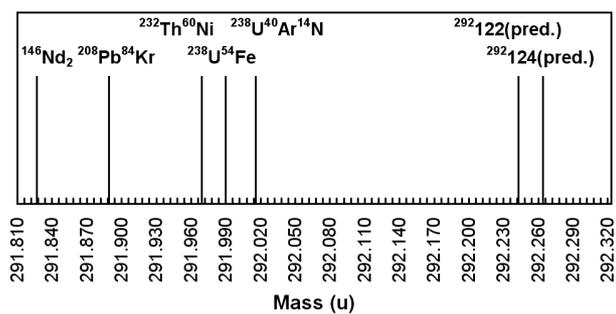

Figure 1.

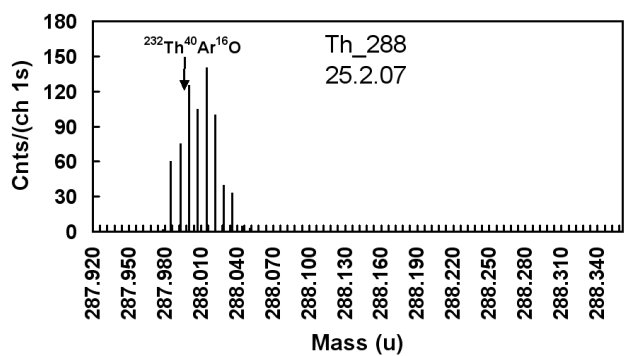

Figure 2.

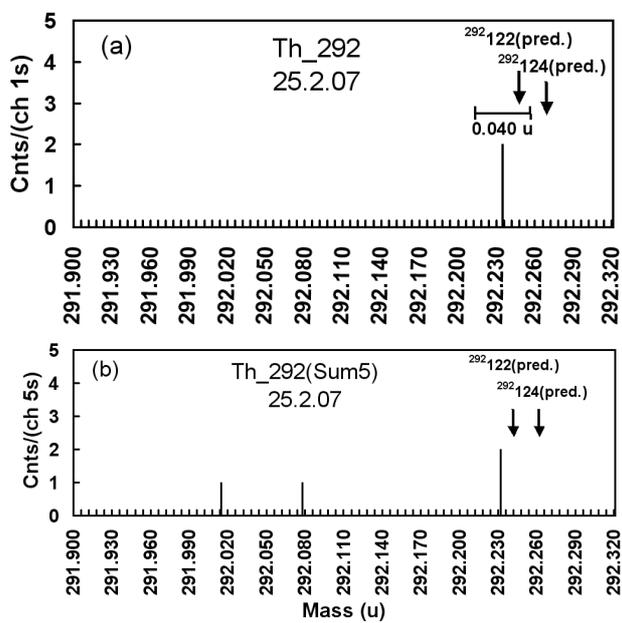

Figure 3.

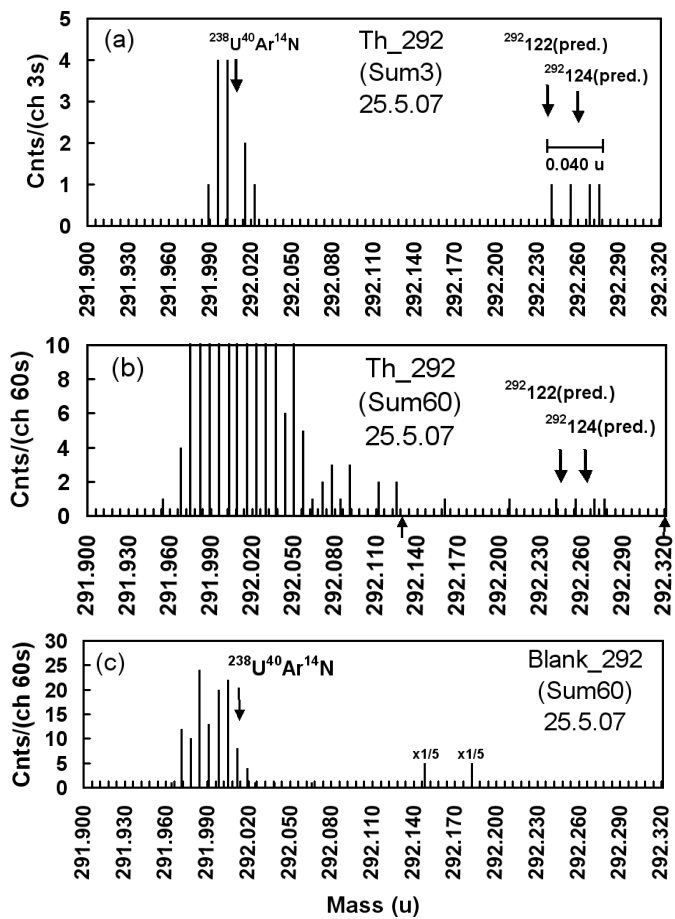

Figure 4.

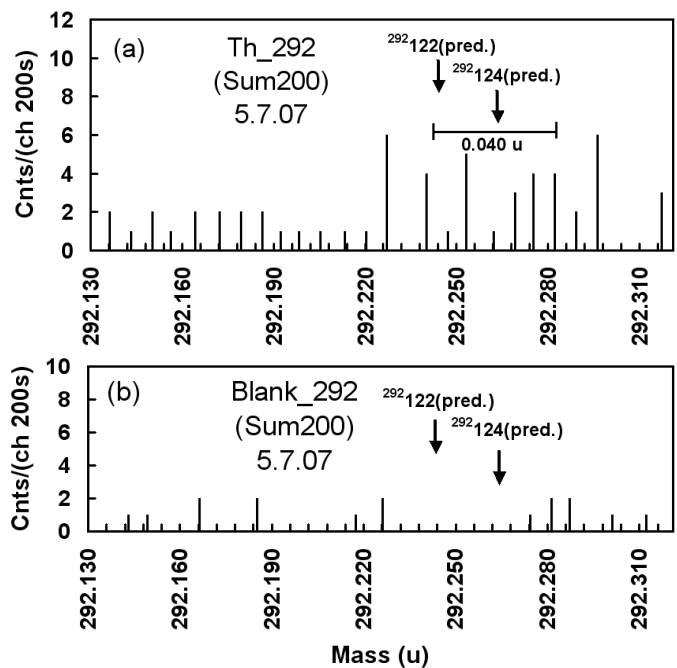

Figure 5.